\documentclass[aps,prl,floats,preprintnumbers,twocolumn,nofootinbib]{revtex4}

\usepackage[dvips]{color,graphicx}
\usepackage{mathrsfs}
\usepackage{amssymb}
\usepackage{amsmath}
\usepackage{feynmp}
\usepackage{slashed}
\usepackage{latexsym}
\usepackage{graphicx}
\usepackage{verbatim}
\usepackage[normalem]{ulem}

\newcommand{\be}{\begin{equation}}
\newcommand{\ee}{\end{equation}}
\newcommand{\ba}{\begin{array}}
\newcommand{\ea}{\end{array}}
\newcommand{\bea}{\begin{eqnarray}}
\newcommand{\eea}{\end{eqnarray}}
\newcommand{\balg}{\begin{align}}
\newcommand{\ealg}{\end{align}}
\newcommand{\bit}{\begin{itemize}}
\newcommand{\eit}{\end{itemize}}
\newcommand{\trm}[1]{\textrm{#1}}

\newcommand{\Mpc}{\trm{\Mpc}}
\newcommand{\yr}{\trm{\yr}}
\newcommand{\eV}{\trm{\eV}}

\begin{document}

\preprint{NUHEP-TH/14-08}

\title{Solar Neutrinos and the Decaying Neutrino Hypothesis}

\author{Jeffrey M. Berryman}
\affiliation{Northwestern University, Department of Physics \& Astronomy, 2145 Sheridan Road, Evanston, IL~60208, USA}

\author{Andr\'e de Gouv\^ea}
\affiliation{Northwestern University, Department of Physics \& Astronomy, 2145 Sheridan Road, Evanston, IL~60208, USA}

\author{Daniel Hern\'{a}ndez}
\affiliation{Northwestern University, Department of Physics \& Astronomy, 2145 Sheridan Road, Evanston, IL~60208, USA}


\begin{abstract}
We explore, mostly using data from solar neutrino experiments, the hypothesis that the neutrino mass eigenstates are unstable. We find that, by combining $^8$B solar neutrino data with those on $^7$Be and lower-energy solar neutrinos, one obtains a mostly model-independent bound on both the $\nu_1$ and $\nu_2$ lifetimes. We comment on whether a nonzero neutrino decay width can improve the compatibility of the solar neutrino data with the massive neutrino hypothesis. 
\end{abstract}

\maketitle

\setcounter{equation}{0}
\setcounter{footnote}{0}

The discovery of distinct nonzero neutrino masses and nontrivial lepton mixing opened the door to several fundamental questions that revolve around the properties of the neutral leptons. Here we concentrate on what, experimentally and model-independently, is known about the neutrino lifetime. 

In the absence of interactions and degrees of freedom beyond those of the Standard Model, the two heaviest neutrinos -- $\nu_2$ and $\nu_3$  ($\nu_1$ and $\nu_2$) in the case of the so-called normal (inverted) neutrino mass hierarchy \cite{Agashe:2014kda} -- are unstable, decaying into lighter neutrinos and photons ($\nu_i\to\nu_j\nu_k\nu_l$ or $\nu_i\to\nu_j+\gamma$, where $i,j,k,l=1,2,3$). The associated lifetimes, given the tiny neutrino masses, are longer than $10^{37}$~years -- much longer than the age of the universe. The presence of new interactions, degrees of freedom, etc., can, of course, change the picture dramatically. 

Experimental bounds on the lifetimes of the neutrinos are much shorter than those expected from the Standard Model minimally augmented to include nonzero neutrino masses. 
Consulting `The Review of Particle Physics' \cite{Agashe:2014kda}, one encounters different bounds that span almost twenty orders of magnitude. Bounds on the neutrino magnetic moment, for example, translate into bounds on radiative neutrino decays \cite{Broggini:2012df}. Results from cosmic surveys sensitive to the expansion rate of the universe at different epochs are consistent with the existence of around three independent neutrino states, naively indicating that these do not decay within the span of billions of years. In order to translate measurements of the expansion rate of the universe into a bound on the neutrino lifetime, however, one must consider the nature of the neutrino decay process, since the daughters of the putative decay also contribute to the expansion rate of the universe and could, in principle, mimic the contributions of their parents \cite{Beacom:2004yd,Hannestad:2005ex}. The observation of the effects of nonzero neutrino masses in cosmic surveys might change the picture significantly \cite{Serpico:2007pt}. 

Model-independent bounds exist from experiments where the number of neutrinos produced in the source can be compared to the number of neutrinos detected some distance away. These include all neutrino oscillation experiments. Given a baseline $L$ and a ``beam'' energy $E$, one expects to be sensitive to a neutrino decay width $\Gamma_i$ for $\nu_i$ with mass $m_i$ such that 
\begin{equation}
\Gamma_i m_i\frac{L}{E}\equiv d_i\frac{L}{E}=5.07\left(\frac{d_i}{\rm eV^2}\right)\left(\frac{L}{\rm km}\right)\left(\frac{\rm GeV}{E}\right)\gtrsim 1. 
\label{eq:estimate}
\end{equation}
Here, for convenience, we define $d_i\equiv\Gamma_i m_i$, which has dimensions of energy-squared, for two reasons. On one hand, all bounds discussed here are sensitive to $d_i$: it is not possible to disentangle the neutrino mass from its decay width, both being unknown. On the other hand,  $d_i$, measured in eV$^2$, can be easily and directly compared to the neutrino mass-squared differences that are measured in neutrino oscillation experiments and ``compete'' with the decay effects. For conversion purposes, $d_i=10^{-11}$~eV$^2$ translates into a lifetime $\tau_i=70$~$\mu$s for a neutrino with mass $m_i=1$~eV.   

Using Eq.~(\ref{eq:estimate}), it is easy to naively estimate that long-baseline accelerator experiments like MINOS, T2K, and No$\nu$A, with $L/E\sim 10^3$~km/GeV, are sensitive to $d_i\gtrsim 10^{-4}$~eV$^2$, atmospheric neutrino experiments like SuperKamiokande, with $L/E\lesssim 10^5$~km/GeV, are sensitive to $d_i\gtrsim 10^{-6}$~eV$^2$, and the KamLAND reactor neutrino experiment, with $L/E\lesssim 2\times 10^4$~km/GeV, is sensitive to $d_i\gtrsim 10^{-5}$~eV$^2$. Detailed analyses of atmospheric and MINOS data, for example, translate into $d_3\lesssim10^{-5}$~eV$^2$ \cite{GonzalezGarcia:2008ru} and $d_3<1.2\times 10^{-4}$~eV$^2$ \cite{Gomes:2014yua}, respectively, assuming $d_1,d_2\ll d_3$.

Astrophysical neutrinos, when directly observed in Earth-bound detectors, provide significantly more stringent bounds on some of the $d_i$. The observation of neutrinos from Supernova 1987A implies that at least one of the neutrino mass eigenstates made it from the explosion to the Earth and can be translated into $d_i<1.2\times 10^{-21}$~eV$^2$ for at least one $i=1,2,3$ \cite{Frieman:1987as}. A very strong bound on at least one of the $d_i$ can also be derived \cite{Beacom:2002vi,Meloni:2006gv,Baerwald:2012kc,Pakvasa:2012db,Dorame:2013lka,Fu:2014gja} from the current and future observations of ultra-high-energy neutrinos using the IceCube detector \cite{Aartsen:2013bka}. 

Solar neutrinos have $L/E\sim 10^{11}$~km/GeV and hence are sensitive to $d_i\gtrsim 10^{-12}$~eV$^2$. The authors of \cite{Beacom:2002cb} were the first to point out that the $^8$B solar neutrino data translate into a very robust bound on $d_2\lesssim10^{-11}$~eV$^2$, mostly independent from $d_1$ and $d_3$. In this letter, we revisit the impact of decaying neutrinos on solar neutrino data. Since the publication of \cite{Beacom:2002cb}, our understanding of solar neutrinos and neutrino properties improved significantly. More and more precise KamLAND data not only confirmed the neutrino oscillation interpretation of solar neutrino data, but also provided a precision measurement of the ``solar'' mass-squared difference, $\Delta m^2_{12}\equiv m_2^2-m_1^2$, and a good independent measurement of the ``solar'' mixing angle $\theta_{12}$ \cite{Gando:2010aa}. Borexino data allow a precision measurement of $^7$Be solar neutrinos, and a clean measurement of the $pp$ solar neutrinos \cite{Bellini:2011rx}. Finally, recent reactor \cite{An:2013zwz,Ahn:2012nd,Abe:2012tg} and accelerator data \cite{Abe:2013hdq} have measured the ``reactor angle'' $\theta_{13}$, revealing that it is nonzero but quite small, $\sin^2\theta_{13}\sim 0.02$. We will argue that all this information allows one to place, almost model-independently, bounds on both $d_1$ and $d_2$ from solar neutrino data. These results, when combined with results from atmospheric neutrinos, allow one to unambiguously place bounds on all three $d_i$, $i=1,2,3$, which are robust, mostly model independent, and do not depend on the values of the neutrino masses or the neutrino mass hierarchy.  

We will show, {\it a posteriori}, that decay effects are negligible for the $L/E$ values probed by the KamLAND experiment. This implies that the oscillation results obtained from KamLAND apply even if the neutrinos have a finite lifetime, including the fact that $\Delta m^2_{12}\sim 10^{-4}$~eV$^2$ and $\sin^22\theta_{12}\sim 0.8$. This in turn implies that neutrino oscillations from the core to the edge of the Sun, to a very good approximation, satisfy the adiabatic approximation. Ignoring the (very small) day--night effect but taking into account that the different neutrinos can decay into final states not accessible to the different solar neutrino detectors, the probability $P_{e\alpha}$ that a neutrino with energy $E$ born in the Sun as a $\nu_e$ is detected as a $\nu_{\alpha}$, $\alpha=e,\mu,\tau$ one astronomical unit $L_{\odot}$ away from the Sun is
\begin{equation}
P_{e\alpha}(E)\simeq \sum_{i=1,2,3}p_{e i}(E)|U_{\alpha i}|^2 e^{-d_i L_{\odot}/E}, 
\label{eq:Pea}
\end{equation}
where $U_{\alpha i}$, $i=1,2,3$, are the elements of the neutrino mixing matrix, while $p_{e i}(E)$ are the probabilities that the neutrino exits the Sun as $\nu_i$ neutrino mass eigenstates. Strictly speaking, Eq.~(\ref{eq:Pea}) is a good approximation when $d_i R_{\odot}/E\ll 1$ where $R_{\odot}$ is the average solar radius. We will show that this is indeed the case for $d_1$ and $d_2$, and we argue in the next paragraph that solar data are not sensitive to $d_3$ effects.  

Given that $|\Delta m^2_{13}|\sim 2\times 10^{-3}$~eV$^2$ -- even if one includes nonzero neutrino decay widths \cite{GonzalezGarcia:2008ru,Gomes:2014yua} -- $p_{e3}(E)\simeq |U_{e3}|^2$ for all relevant solar neutrino energies, $E\in[100~{\rm keV},20~{\rm MeV}]$. Since $|U_{e3}|^2=\sin^2\theta_{13}\simeq 0.02$ is small, given the precision of the solar neutrino data, $d_3$ related effects are irrelevant. In other words, the solar data are consistent with all $d_3$ values. We anticipate that $d_3\neq 0$ effects impact only very modestly the constraints on the other oscillation and decay parameters. Henceforth, we ignore $\theta_{13}$ effects -- we formally set it to zero -- and treat solar neutrino oscillations as if there were only two neutrinos, $\nu_e$ and $\nu_a$ ($a$ for active).     

\begin{figure}
\includegraphics[width=0.45\textwidth]{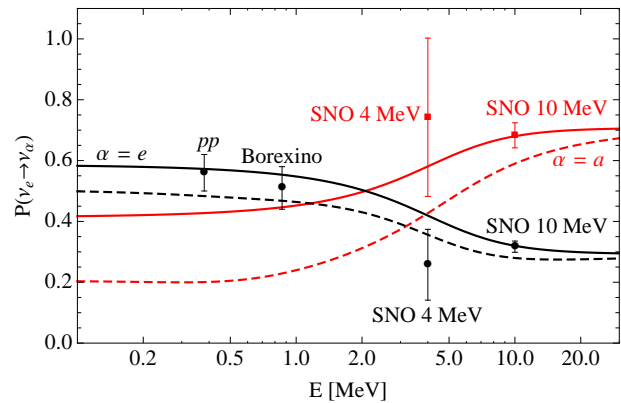}
\caption{$P_{ee}$ (black) and $P_{ea}$ (red) as a function of the solar neutrino energy, for $\sin^2\theta_{12}=0.29$, $\Delta m^2_{12}=7.5\times 10^{-5}$~eV$^2$, $d_1=0$ and $d_2=0$ (solid) or $d_2=2\times 10^{-12}$~eV$^2$ (dashed). Also depicted are the data points used to estimate the allowed values of $d_1$ and $d_2$, including one sigma error bars. See text for details.}
\label{fig:ps}
\end{figure}

At high solar neutrino energies, $E\gtrsim 5$~MeV, $p_{e2}\sim 1$, $p_{e1}\sim 0$, such that $P_{ee}\sim \sin^2\theta_{12}e^{-d_2L_{\odot}/E}$ and $P_{ea}\sim \cos^2\theta_{12}e^{-d_2L_{\odot}/E}$. $^8$B solar neutrino data are hence very sensitive to $d_2$ but have little sensitivity to $d_1$ \cite{Beacom:2002cb}. The most recent solar data from SNO \cite{Aharmim:2011vm} indicate a $P_{ee}$ that decreases slowly as the neutrino energy decreases (as opposed to increasing, as predicted by the standard scenario, $d_1=d_2=0$), a fact that is consistent with a judicious choice of $d_2$. For illustrative purposes, Fig.~\ref{fig:ps} depicts $P_{ee}$ and $P_{ea}$ as a function of $E$, for $\sin^2\theta_{12}=0.29$, $\Delta m^2_{12}=7.5\times 10^{-5}$~eV$^2$, $d_1=0$, and $d_2=0$ or $d_2=2\times 10^{-12}$~eV$^2$.

At low solar neutrino energies, $E\lesssim 1$~MeV, solar neutrino oscillations are well approximated by simple, averaged-out vacuum oscillations such that $p_{e1}\sim \cos^2\theta_{12}$, $p_{e2}\sim \sin^2\theta_{12}$, and $P_{ee}\sim  \cos^4\theta_{12}e^{-d_1L_{\odot}/E}+\sin^4\theta_{12}e^{-d_2L_{\odot}/E}$ and $P_{ea}\sim \sin^2\theta_{12}\cos^2\theta_{12}(e^{-d_2L_{\odot}/E}+e^{-d_1L_{\odot}/E})$. $^7$Be and $pp$ solar neutrino measurements are hence sensitive to both $d_1$ and $d_2$. In isolation, the low-energy solar neutrino data can be used to place a bound on either $d_1$ or $d_2$, but not both. This is easy to see: the data are consistent with $e^{-d_1L_{\odot}/E}\to 0$ or $e^{-d_2L_{\odot}/E}\to 0$ as long as one judiciously chooses $\sin^2\theta_{12}$. For example, in the limit, say, $e^{-d_1L_{\odot}/E}\to 0$, $P_{ee}\sim \sin^4\theta_{12}$ and $P_{ea}\simeq \sin^2\theta_{12}\cos^2\theta_{12}$ can be made to fit the data, roughly, $P_{ee}\sim 0.55$ \cite{Bellini:2011rx}, by choosing $\sin^2\theta_{12}=0.75$ (in the ``dark side'' \cite{de Gouvea:2000cq}), which is consistent with data from KamLAND \cite{Gando:2010aa}. This possibility, however, is ruled out by $^8$B data, which ``require'' $\sin^2\theta_{12}\sim 0.3$. 

In summary, combined low and high energy solar neutrino data allow one to place nontrivial bounds on both $d_1$ and $d_2$, i.e., the possibility that either $e^{-d_1L_{\odot}/E}\to 0$ or $e^{-d_2L_{\odot}/E}\to 0$ is ruled out. $d_2$ is mostly constrained by the $^8$B data, while $d_1$ is mostly constrained by the $^7$Be and $pp$ data. Given the order-of-magnitude difference between the neutrino energies, we anticipate the $d_1$ bound to be, roughly, an order of magnitude stronger than the $d_2$ bound. 

In order to estimate the upper bounds on $d_1$ and $d_2$, we perform a simple $\chi^2$ fit to $\sin^2\theta_{12}, d_1, d_2$, fixing $\Delta m^2_{12}=7.5\times 10^{-5}$~eV$^2$, the best fit from KamLAND, and setting $\sin^2\theta_{13}=0$, as discussed earlier. Given that KamLAND provides the dominant contribution to the measurement of $\Delta m^2_{12}$, this is a very reasonable approximation. Since in the decaying-neutrinos scenario $P_{ee}+P_{ea}\le1$, we need to consider separately the information on the electron and the ``active'' neutrino components of the solar neutrino flux. In detail, we include the following experimental information, depicted in Fig.~\ref{fig:ps}:
\begin{itemize}
\item $P_{ee}=0.56\pm 0.06$ for $E=380$~keV, as extracted from a combined fit to Borexino and low-energy neutrino data \cite{Bellini:2011rx}. This analysis is performed, effectively, by using Borexino data in order to establish the oscillated $^7$Be neutrino flux and hence extract the $pp$ neutrino flux from other data. This procedure depends only weakly on the hypothesis that $P_{ee}+P_{ea}=1$. This result is consistent with Borexino's recent independent measurement of the low energy solar neutrino flux \cite{Bellini:2014uqa}.
\item $s=P_{ee}+rP_{ea}=0.62\pm 0.05$ for $E=862$~keV from the Borexino data \cite{Bellini:2011rx}.  $r=0.22$ is the ratio of the $\nu_e + e$ to the $\nu_a + e$ elastic scattering cross-sections at $^7$Be neutrino energies. 
\item SNO performed a detailed measurement of $P_{ee}$ as a function of energy \cite{Aharmim:2011vm}. We choose $P_{ee}$ values at $E=4~{\rm MeV}$ and $E=10~{\rm MeV}$, $P_{ee}=0.26\pm 0.12$ and $P_{ee}=0.32\pm 0.02$, respectively, as representatives of the SNO data. These points are chosen in order to both capture the statistical power of the SNO experiment and to include some of the shape information. A proper treatment of the SNO data, including all different observables, correlations, etc., can only be handled by the Collaboration itself. We verify that, in the case $d_1=d_2=0$, our extracted best fit value for $\sin^2\theta_{12}$ and the associated one sigma error bar are in good agreement with the most recent global analyses of neutrino data \cite{Gonzalez-Garcia:2014bfa}. 
\item The SNO experiment is also sensitive to the presence of a $\nu_a$ flux from the Sun thanks to its neutral current and $\nu+e$ elastic scattering measurements. It is, therefore, possible to measure $P_{ee}$ and $P_{ea}$ as a function of energy with SNO data (see, for example, \cite{Ahmad:2002jz}). Ref.~\cite{Aharmim:2011vm}, however, does not discuss the independent extraction of $P_{ea}$ from the data, replacing it instead by $1-P_{ee}$. Here we estimate the extracted value of $P_{ea}$ from SNO data as follows. We define the central value using $P_{ea}=1-P_{ee}$ while fixing the one-sigma error bar on $P_{ea}$ as that on $P_{ee}$, multiplied by $\sqrt{5}$. The factor of $5$ is very close to the ratio of the elastic $\nu_e+e$ cross-section to that for $\nu_a+e$ at $^8$B neutrino energies and agrees with the relative uncertainties for the electron and active neutrino fluxes measured by SNO in  \cite{Ahmad:2002jz}.
\end{itemize}
We note that SuperKamiokande also measures the neutrino flux using elastic neutrino--electron scattering (see, e.g., \cite{Abe:2010hy}). We do not include data from SuperKamiokande in our simplified fit as they mostly contributes to the measurement of $P_{ea}$ -- which we can only estimate here -- and have a higher energy threshold than SNO data.

\begin{figure}
\includegraphics[width=0.45\textwidth]{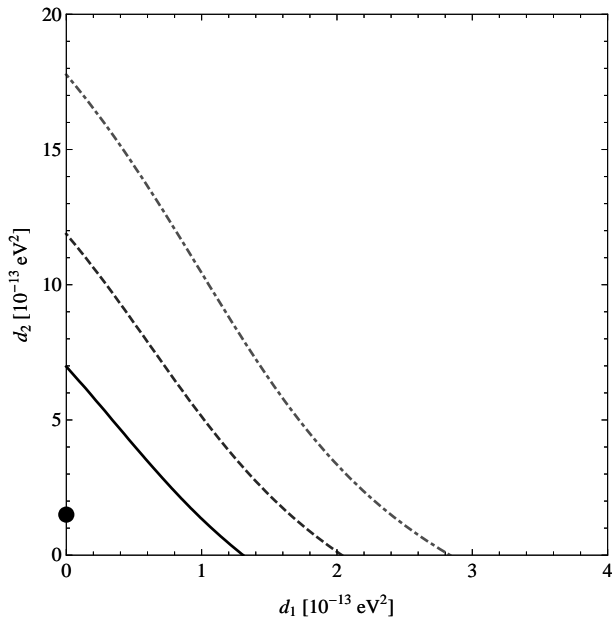}
\caption{Allowed values of $d_1$ and $d_2$ at the one, two and three sigma. The best fit point $d_1=3.4\times 10^{-19}~{\rm eV}^2, d_2=1.6\times 10^{-13}$~eV$^2$ is indicated by a dot.}
\label{fig:ds}
\end{figure}

Fig.~\ref{fig:ds} depicts the result of our fit in the $d_1\times d_2$-plane, obtained after marginalizing over $\sin^2\theta_{12}$. The best fit point is $d_1=3.4\times 10^{-19}~{\rm eV}^2, d_2=1.6\times 10^{-13}$~eV$^2$ and the hypothesis $d_1=d_2=0$ fits the data quite well. At the two-sigma confidence level, $d_1<1.6\times 10^{-13}$~eV$^2$ and $d_2<9.3\times 10^{-13}$~eV$^2$, in agreement with the naive estimates discussed above. The constraints above justify the approximations that led to  Eq.~(\ref{eq:Pea}), especially $d_{1,2}R_{\odot}/E\ll1$ for all solar neutrino energies. Our result indicates that the neutrino decay hypothesis does not allow for a fit to the solar data that is significantly better than the standard ``large mixing angle solution,'' mostly due to the low energy $^7$Be and $pp$ neutrino measurements. We emphasize, however, that a detailed analysis of all solar neutrino data including the neutrino decay hypothesis is best left to the experimental collaborations, and that a reanalysis of the SNO data -- one that treats both $P_{ee}(E)$ and $P_{ea}(E)$ as independent functions  -- as a function of energy is required. We hope our results encourage the pursuit of such an analysis.

In summary, we have argued that the solar neutrino data, combined with reactor data, allow one to place mostly model-independent bounds on the lifetimes of $\nu_1$ and $\nu_2$. Using a subset of the solar neutrino data and the data from KamLAND, we estimate that $d_1<1.6\times 10^{-13}$~eV$^2$ and $d_2<9.3\times 10^{-13}$~eV$^2$ at the two-sigma confidence level. A complete analysis would reveal exactly where these bounds lie. As a ``by-product'' of our analysis, the atmospheric neutrino bound discussed  in \cite{GonzalezGarcia:2008ru} applies, robustly, to $\nu_3$: $d_3\lesssim 10^{-5}$~eV$^2$. Along with solar and atmospheric neutrino data, the only other robust bound comes from SN1987A which translates, as discussed earlier, into $d_i<1.2\times 10^{-21}$ for one of the three mass-eigenstates, most likely $\nu_1$ or $\nu_2$. 

The bounds are `mostly model-independent' in the following sense. They are independent from the values of the neutrino masses themselves, and apply for both mass hierarchies. No assumption is made regarding the nature of the neutrino -- Majorana or Dirac -- or of the daughter particles into which the neutrinos would be decaying. We are assuming, however, that if the decay-daughters were to consist of lighter active neutrinos, these would not leave a significant imprint in the detectors under consideration, i.e., they don't ``look'' like the parent neutrinos. This is a modest assumption. Daughter neutrinos from neutrino decay have, necessarily, less energy than their parents, and only those that decay along the flight-path of the parent make it to the detector. We also do not allow for the possibility, recently discussed in a more generic context \cite{Berryman:2014yoa}, that the neutrino decay hypothesis translates into more mixing parameters, i.e., that the neutrino mass and decay eigenstates are not the same.

\section*{Acknowledgements}

We thank Roberto Oliveira for useful discussions and collaboration during the early stages of this work. DH and AdG thank the Kavli Institute for Theoretical Physics in Santa Barbara, where part of this work was completed, for its hospitality.  DH benefitted from fruitful discussions with Hisakazu Minakata, Sandip Pakvasa, Tom Weiler, and Walter Winter. The work of AdG and DH was supported in part by the National Science Foundation under Grant No. NSF PHY11-25915. This work is sponsored in part by the DOE grant \#DE-FG02-91ER40684.

\end{document}